\documentclass[prl,twocolumn,showpacs,amstext,amsxtra,amssymb,latexsym,bbm,superscriptaddress]{revtex4}
\usepackage{graphicx,epsf,subfigure}
\usepackage{amsmath,float}
\usepackage{amsfonts}
\usepackage{amsthm}
\usepackage{amssymb}
\usepackage{multirow}
\usepackage{color}

\newcommand{\K}{{\boldsymbol K}}
\renewcommand{\k}{{\boldsymbol k}}

\newcommand{\be}{\begin{equation}}
\newcommand{\ee}{\end{equation}}

\renewcommand{\a}{{\boldsymbol a}}

\newcommand{\q}{{\boldsymbol q}}

\newcommand{\ep}{\epsilon}

\newcommand{\vf}{v}

\begin{document}

\title{From dia- to paramagnetic orbital susceptibility of Dirac cones}
 \author{A. Raoux}
 \affiliation{Laboratoire de Physique des Solides, CNRS UMR 8502, Univ. Paris-Sud, F-91405 Orsay, France}
 \author{M. Morigi}
 \affiliation{Laboratoire de Physique des Solides, CNRS UMR 8502, Univ. Paris-Sud, F-91405 Orsay, France}
 \author{J.-N. Fuchs}
 \affiliation{Laboratoire de Physique Th\'eorique de la Mati\`ere Condens\'ee, CNRS UMR 7600, Universit\'e Pierre et Marie Curie, 4 place Jussieu, F-75252 Paris, France}
 \affiliation{Laboratoire de Physique des Solides, CNRS UMR 8502, Univ. Paris-Sud, F-91405 Orsay, France}
 \author{F. Pi\'echon}
 \affiliation{Laboratoire de Physique des Solides, CNRS UMR 8502, Univ. Paris-Sud, F-91405 Orsay, France}
 \author{G. Montambaux}
 \affiliation{Laboratoire de Physique des Solides, CNRS UMR 8502, Univ. Paris-Sud, F-91405 Orsay, France}
 \date{\today}

\begin{abstract}
We study the orbital susceptibility of coupled energy bands with a pair of Dirac points, as in graphene. We show that different systems having the {\it same} zero-field energy spectrum exhibit strong differences in their orbital magnetic response at zero energy, ranging from diamagnetism (graphene) to paramagnetism (dice lattice). A lattice model is introduced which interpolates continuously between these two limits. This striking behavior is related to a Berry phase varying continuously from $\pi$ to $0$. These predictions could be tested with cold atoms in an optical lattice.
\end{abstract}
\maketitle

\begin{figure}[h!]
  \centering
  \subfigure[]{\includegraphics[width=4.5cm]{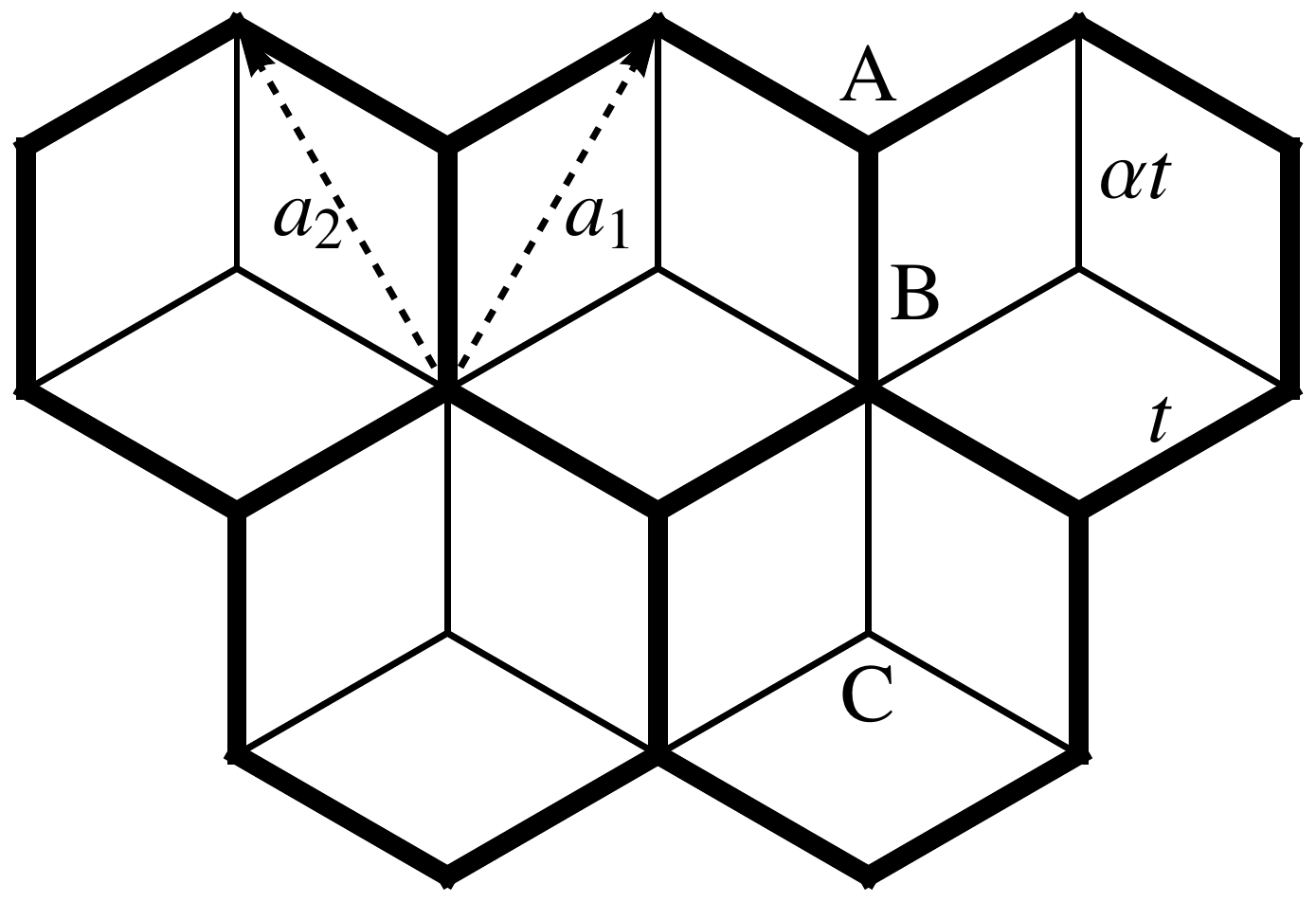}}
   \subfigure[]{\includegraphics[width=3.5cm]{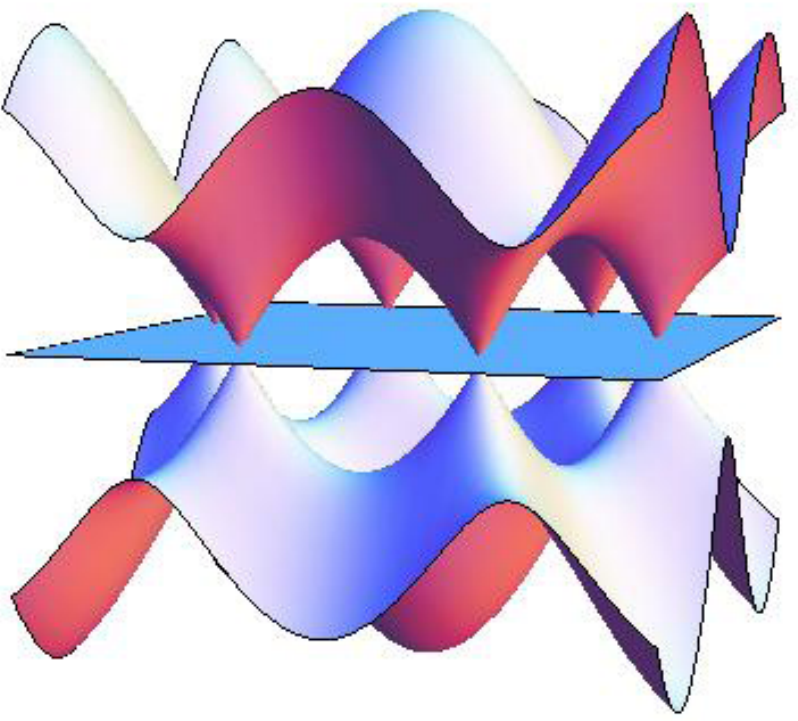}}
  \caption{$\alpha$-$\mathcal{T}_3$ model. (a) $\mathcal{T}_3$ lattice. Thick links: nearest neighbours hoppings $t$ between $A$ and $B$ sites forming a honeycomb lattice. Thin links: additional hoppings $\alpha t$ connecting $C$ to $B$ sites. Varying $\alpha$ interpolates between the honeycomb lattice ($\alpha=0$) and the dice lattice ($\alpha=1$).  (b) Zero field energy spectrum as a function of the wavevector $\k$ for all $\alpha$.
}
		\label{fig:lattice}
\end{figure}

\begin{figure}
\begin{center}
  \subfigure[]{\includegraphics[width=5.1cm]{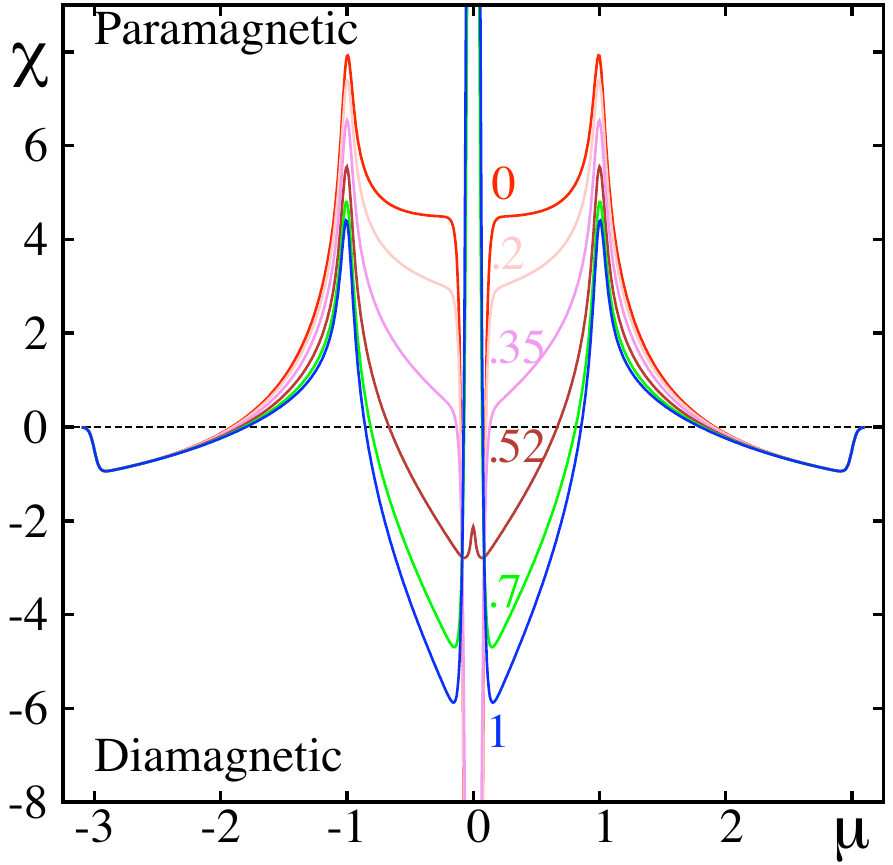}}
   \subfigure[]{\includegraphics[width=3.4cm]{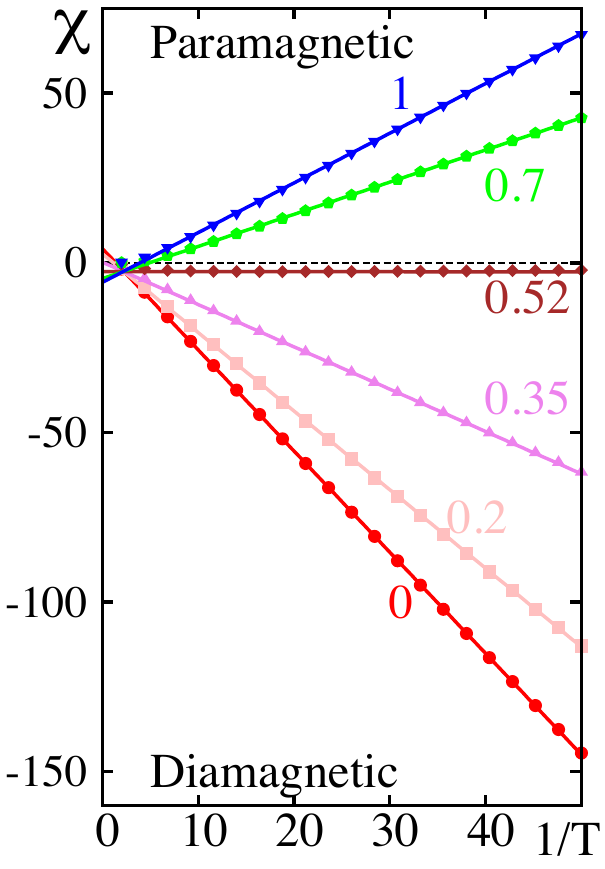}}
 \caption{ (a) Numerically obtained orbital susceptibility $\chi$ (in units of the Landau band edge value $|\chi_L|=\frac{1}{16\pi }\frac{e^2 t a^2}{\hbar^2}$) as a function of the chemical potential $\mu$ (in units of $t$) in the whole band for various $\alpha$ as indicated and for a temperature $T=0.02t$. (b) Susceptibility peak at zero chemical potential as a function of the inverse temperature for various $\alpha$ (same units as in (a)). The slope is well fitted by $-3\zeta_\gamma(2)/\zeta(2)$, in agreement with Eq.~(\ref{chichi0}).
 }
 \label{fig:chi-de-E-alpha}
\end{center}
\end{figure}

{\it Introduction - } Among the fascinating electronic properties of graphene, orbital magnetism is certainly one of the least studied experimentally \cite{RMP}. This is due to the difficulty of measuring the magnetic response of such a thin solid \cite{Sepioni}. Theoretically, it has long been known that undoped graphene should be strongly diamagnetic \cite{McClure}. This behavior is attributed to the response of Dirac-Weyl fermions which properly describe the low energy electronic properties of graphene. Here, we show that another simple system also featuring Dirac-Weyl fermions, the so-called dice (or ${\cal T}_3$) lattice \cite{Sutherland,Vidal}, surprisingly presents a huge {\it paramagnetic} response in a magnetic field. These two systems have the {\it same} zero field energy spectrum (see Fig.~\ref{fig:lattice}b) but exhibit opposite magnetic behaviors (see Fig.~\ref{fig:chi-de-E-alpha}). To shed light on this striking difference, we propose and study a lattice model which interpolates between these two extreme limits.

The honeycomb (graphene) and the dice lattices are examples of electronic systems featuring coupled energy bands for which the Landau-Peierls (or single-band) approach \cite{LP} fails to obtain the orbital susceptibility $\chi$. Here the band of itinerant electrons is made of two sub-bands touching at two Dirac points. In the case of graphene, by considering the vicinity of the Dirac points where the spectrum is linear, McClure showed that the magnetic field spectrum exhibits peculiar Landau levels (LLs) \cite{McClure}. Neglecting the contribution of the rest of the band, he could derive a diamagnetic peak precisely at the Dirac point, i.e. for zero doping, where Landau-Peierls would predict a vanishing susceptibility. Later, Fukuyama developed a linear response formalism that takes inter-band effects into account \cite{Fukuyama}. This formalism has recently been applied to the tight-binding model of graphene to calculate $\chi(\mu)$ where the chemical potential $\mu$ varies in the entire band \cite{Stauber}.
The divergence of the susceptibility at $\mu=0$ appears as a $\delta(\mu)$ peak, which is the signature of a non analytic behavior of the grand potential as a function of the magnetic field \cite{Sharapov}.

In order to get a better understanding of the fundamental reason for this peculiar behavior,  we introduce and study a modified tight-binding model for spinless electrons hopping on the ${\cal T}_3$ lattice,
which we call $\alpha$-${\cal T}_3$. Starting from the honeycomb lattice with two sites $(A,B)$ per unit cell and a hopping amplitude $t$, the ${\cal T}_3$ lattice is obtained by connecting additional ($C$) sites at the center of each hexagon to the $B$ sites (see Fig.~\ref{fig:lattice}a). The additional hopping amplitude (linking $B$ to $C$) is denoted $\alpha t$. Depending on the real parameter $\alpha$, this model interpolates between graphene ($\alpha=0$) and the dice lattice ($\alpha=1$). 
Its interest is that the zero field spectrum (within a trivial  appropriate normalisation) does \textit{not} depend on $\alpha$ (see Fig.~\ref{fig:lattice}b), while the zero field wave functions and the finite field spectrum present a continuous evolution which may be described by an $\alpha$-dependent Berry phase. Then, we compute the low-energy LLs, from which we obtain the low field dependence of the magnetization. By considering different temperature limits, we explicitly show that the magnetic response \textit{continuously evolves from a diamagnetic to a paramagnetic behavior}, when increasing $\alpha$. We finally present simple arguments to explain this spectacular change of the orbital response, backup our low energy analytical results by numerical calculations on the full tight-binding model and suggest an experimental realization.

{\it The $\alpha$-${\cal T}_3$ model - }
We first introduce a convenient parametrization of $\alpha$ with the angle $\varphi$ such that $\tan \varphi\equiv \alpha$. 
 Due to the three sites per unit cell, the Bloch Hamiltonian has a $3 \times 3$ structure and reads (after rescaling the energy by $\cos \varphi$ \cite{rescaling}):
\be H(\k) =
  \left(
          \begin{array}{ccc}
            0 &  f_\k \cos \varphi & 0 \\
            f_\k^* \cos \varphi   & 0 & f_\k \sin \varphi  \\
            0 &  f_\k^* \sin \varphi & 0 \\
\end{array}
\right) \, ,
\ee
where $f_\k= - t(1+ e^{-i \k . \a_1} + e^{-i \k . \a_2}) $, $\a_1=a(\sqrt{3}/2,3/2)$ and $\a_2=a(-\sqrt{3}/2,3/2)$ are Bravais lattice vectors (Fig.~\ref{fig:lattice}a), $a$ is the inter-site distance and the wavevector $\k=(k_x,k_y)$. The model has a $\alpha \to 1/\alpha$ duality and we therefore restrict ourselves to $\alpha \in [0,1]$. The corresponding spectrum is \textit{independent of $\alpha$} and consists of three bands each of them carrying 1/3 of the states: a zero-energy flat band $\ep_{\k,0}=0$ and two dispersive bands $\ep_{\k,\lambda}= \lambda |f_\k| $, with the band index $\lambda=\pm$, see Fig.~\ref{fig:lattice}b. The latter are identical to the bands of graphene \cite{RMP} and feature two inequivalent contact points at the corners $\pm \K$ of the hexagonal Brillouin zone, where $f(\pm \K)=0$. 
To distinguish these two contact points, we introduce a valley index $\xi=\pm$. Close to zero energy, linearization near $\xi \K$ gives  $f_\k\simeq \vf (\xi q_x - i q_y)$ where the velocity $\vf \equiv 3 t a /2$, $\q= \k - \xi\K$ and we have set $\hbar \equiv 1$. The low-energy spectrum is therefore $\ep_{\q,0}=0$ and $\ep_{\q,\lambda} =\lambda \vf |\q |$ and electrons behave as massless fermions. The eigenvectors in the whole Brillouin zone read
\be |\psi_{\lambda}\rangle = {1 \over \sqrt{2}} \left(
                                                \begin{array}{c}
                                                   \cos \varphi \, e^{i \theta_\k} \\
                                                 \lambda \\
                                                    \sin \varphi  \, e^{-i \theta_\k} \\
                                                \end{array}
                                              \right) \ , \ |\psi_0\rangle = \left(
                                                \begin{array}{c}
                                                  \sin \varphi \,  e^{i \theta_\k} \\
                                                 0 \\
                                                     -\cos \varphi \,  e^{-i \theta_\k} \\
                                                \end{array}
                                              \right) \ee
where $f_\k=|f_\k|e^{i\theta_\k}$ defines the angle $\theta_\k$ and $|\psi_0\rangle$ corresponds to the zero energy flat band. For a path encircling a single valley, the finite energy bands are characterized by a Berry phase $\phi_{\lambda,\xi}= \xi \pi \cos 2 \varphi$, while the flat band has a Berry phase $\phi_{0,\xi} = -\xi 2\pi \cos 2 \varphi\equiv \xi 4\pi \sin^2 \varphi$ (modulo $2\pi$). Note that $\phi_{0,\xi}+\sum_\lambda \phi_{\lambda,\xi}=0$, $\sum_\xi \phi_{\lambda,\xi}=0$ and $\sum_\xi \phi_{0,\xi}=0$ as it should. It is remarkable that, except for $\alpha=0$ or $1$, the Berry phase $\phi_{\lambda,\xi}$ is different in the two valleys. 
These Berry phases are topological but not $\pi$-quantized.
\medskip

{\it Dirac-Weyl Hamiltonians - }
For $\alpha=0$, the Bloch Hamiltonian becomes block diagonal $H(\k)=\left(\begin{array}{cc}0&f_\k\\f_\k^*&0 \end{array}\right)\oplus 0$. The $\alpha$-${\cal T}_3$ model is simply that of graphene except for the additional zero energy flat band originating from the uncoupled $C$ atoms. As this flat band is inert and does not change the physics whatsoever, we will refer to the $0$-${\cal T}_3$ model as \textit{graphene}, notwithstanding the three sites per unit cell.
Close to each valley, the linearized Hamiltonian can be written in the Dirac-Weyl form $H_\xi = \vf \, (\xi q_x \sigma_x +q_y \sigma_y)\oplus 0$, where $\sigma_x,\sigma_y$ are spin 1/2 Pauli matrices \cite{RMP}.

For $\alpha=1$, the $\alpha$-${\cal T}_3$ model is that of the usual dice lattice \cite{Sutherland}. We will refer to the $1$-${\cal T}_3$ model simply as \textit{dice}. In the vicinity of the contact points, the Hamiltonian may be linearized as $H_\xi= \vf \, (\xi q_x S_x + q_y S_y)$, similar to graphene except that $S_x,S_y$ are now spin $1$ matrices. This low-energy Hamiltonian is that of Dirac-Weyl fermions with pseudo-spin 1 \cite{Grabert}. The $\alpha$-${\cal T}_3$ model therefore provides a continuous interpolation between pseudo-spin $1/2$ ($\alpha=0$) and pseudo-spin $1$ ($\alpha=1$) massless fermions. However, when $\alpha\neq 0,1$, the model involves more than a single pseudo-spin operator.
\begin{figure}
\begin{center}
\subfigure[]{\includegraphics[width=4cm]{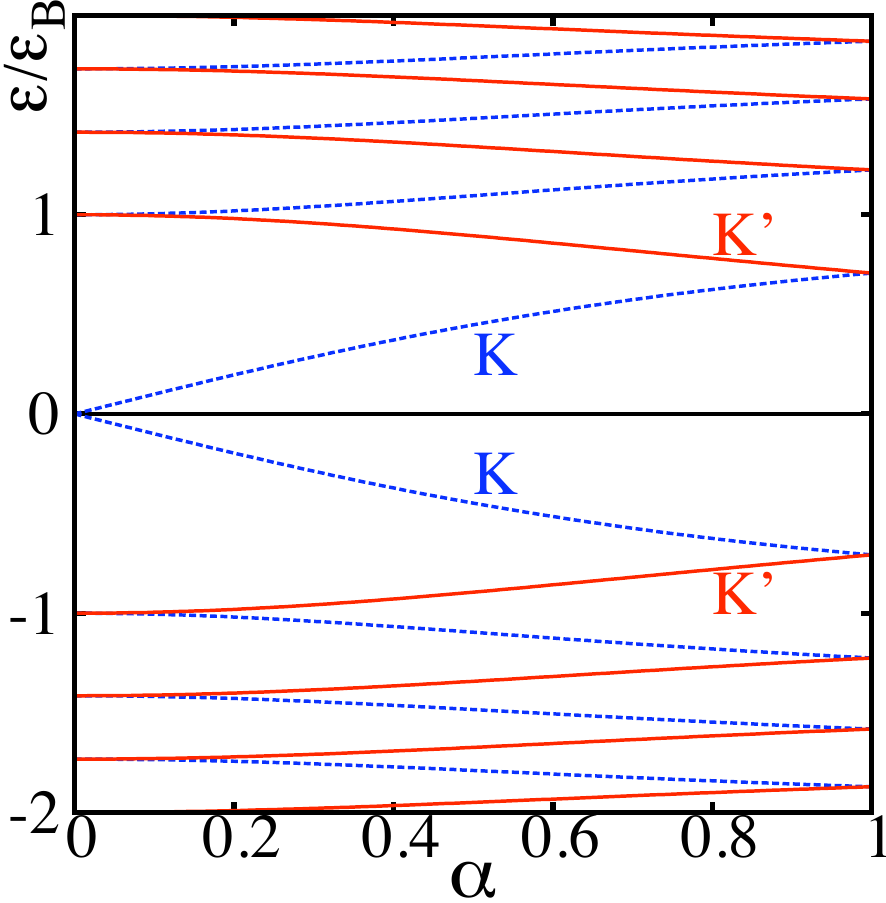}}
\subfigure[]{\includegraphics[width=4.1cm]{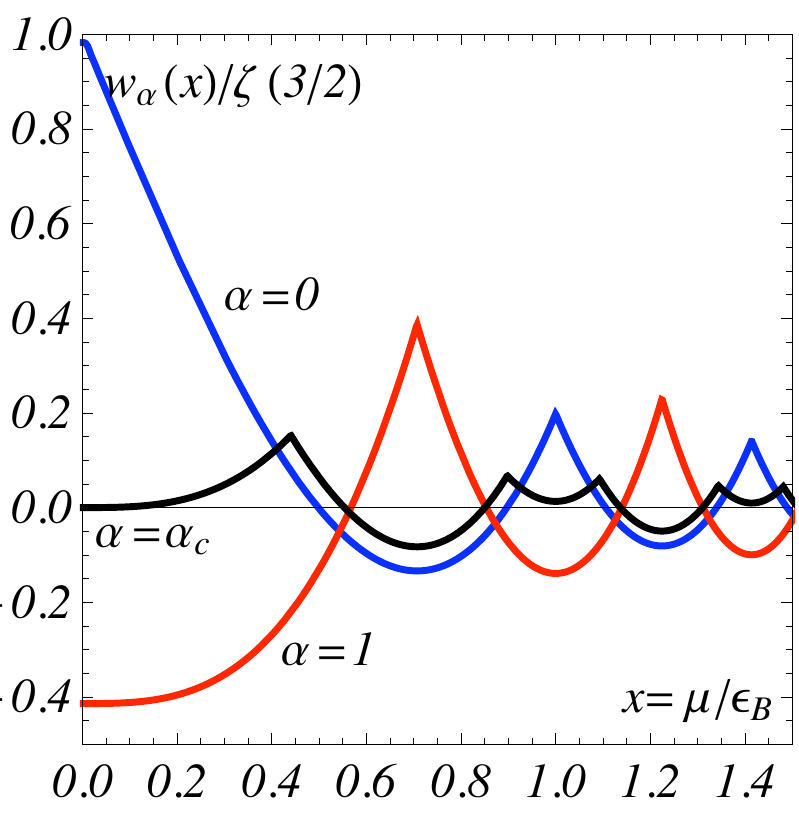}}
 \caption{(a) Landau level spectrum $\ep_n$ (each color corresponding to one valley) near zero energy as a function of the parameter $\alpha$ from graphene ($\alpha=0$) to dice ($\alpha=1$). (b) Function $w_\alpha(x)$ giving the dependence of the grand potential $\delta{{\Omega}}_\alpha$ (at finite magnetic field and zero temperature) on the chemical potential $\mu$, see Eq.~(\ref{GP2}). Blue: graphene ($\alpha=0$). Red: dice lattice ($\alpha=1$). Black: critical case ($\alpha_c= 0.49065$).}
 \label{fig:landau-graphene-to-dice}
\label{fig:walpha}
\end{center}
\end{figure}
\medskip

{\it Landau level spectrum - } We now add a perpendicular magnetic field $B$ to study the evolution of the LL spectrum with the parameter $\alpha$. To do so, we restrict ourselves to the vicinity of the contact points where the zero field spectrum is linear, that is an energy interval $[-W,W]$ where $W$ is an energy cut-off, typically smaller than $t$. Performing the usual Peierls substitution $q_x\pm iq_y \rightarrow \sqrt{2eB} \hat{a}^\dagger/\hat{a}$, that introduces ladder operators such that $[ \hat{a},\hat{a}^\dagger ]=1$, the low-energy Hamiltonian in the $K$ valley becomes:
\be H_+ =\ep_B  \left(
          \begin{array}{ccc}
            0 & \cos \varphi \hat a & 0 \\
            \cos \varphi \hat a^\dagger & 0 &\sin \varphi \hat a \\
            0 & \sin \varphi \hat a^\dagger & 0 \\
          \end{array}
        \right)\, ,\ee
where $\ep_{B} \equiv \vf \sqrt{2 e B}$ is a characteristic magnetic energy. In the other ($K'$) valley, $H_-$ is obtained from $H_+$ by the substitution $\hat{a}\to -\hat{a}^\dagger$. The Landau spectrum in each valley is given by $\ep_{l,\xi}=\pm  \ep_B\sqrt{l+\gamma_\xi}$, where $l \in \mathbb{N}$ is the Landau index and $\gamma_+=\sin^2 \varphi=1-\gamma_-$ is a valley-dependent index shift. The latter is related to the above computed Berry phase $\phi_{\lambda,\xi}$ via the semiclassical relation $\gamma_\xi=1/2-\phi_{\lambda,\xi}/2\pi$, see e.g. \cite{Mikitik,Fuchs}, which is here found to be exact. As $\gamma_\xi$ depends on the valley index, the twofold valley degeneracy is lifted by the magnetic field as soon as $\alpha \neq 0, 1$. In order to treat both valleys at once, it is convenient to relabel the LLs as
\be \ep_n \equiv \pm \ep_B \sqrt{|n+\gamma|} \label{eq:lls}\ee
with $\gamma= \gamma_+$ and a new Landau index $n \in \mathbb{Z}$ that now also takes negative values. When $\alpha\neq 0,1$, each LL $\ep_n$ has a degeneracy $eB/h$ per unit area. LLs are plotted as a function of $\alpha$ in Fig.~\ref{fig:landau-graphene-to-dice}a. For graphene ($\alpha=0$), the Landau spectrum is well-known to be $\ep_n = \pm \ep_B \sqrt{|n|}$, $n \in \mathbb{Z}$, including a zero energy LL \cite{McClure}. For the dice lattice ($\alpha=1$), it is given by $\ep_n = \pm   \ep_B \sqrt{|n+1/2|}$, $n \in \mathbb{Z}$ \cite{Grabert}. For all $\alpha$, in addition to the above discussed LLs, a zero-energy flat band of topological origin exists in an arbitrary magnetic field, carrying 1/3 of the states (for $\alpha=1$, see \cite{Vidal}).
\medskip

{\it Grand potential and magnetization of Dirac cones - } In order to compute the orbital magnetization, we work in the grand canonical statistical ensemble and consider the low energy LL spectrum (\ref{eq:lls}). The grand potential is written as $\Omega(\mu,T)= \int {\cal N}(\ep) f_\mu'(\ep) d\ep$, where ${\cal N}(\ep)$ is the doubly integrated density of states (DoS), measured from the bottom of the spectrum and $f_\mu(\ep)$ is the Fermi-Dirac function with chemical potential $\mu$ and temperature $T=1/\beta$ (with $k_B\equiv 1$) \cite{AM}. It is then convenient to write it as a function of the doubly integrated DoS ${\cal N}_0(\ep)$ measured from zero energy. Neglecting terms which are field independent, we find that the {\it field dependent part } $\delta {\cal N}\equiv {\cal N}(B)-{\cal N}(B=0)$ can be written as $\delta {\cal N} (\ep)=\delta{\cal N}_0(\ep)- \delta{\cal N}_0(-W)$ where $W$ is the energy cut-off (in the low field limit $W/\ep_B\to \infty$) \cite{SM}. The DoS (per unit area) in the vicinity of the contact points is given by
\be \nu(\ep,B)={e B \over h}  \sum_{n, \pm}   \delta\left(\ep
 \pm \ep_{B} \sqrt{|n+ \gamma|}\right)      \ \ , \ee
including the valley degeneracy. The contribution of the flat band, which is field independent, has been excluded. The Poisson formula leads to the  Fourier decomposition of this DoS:
\be \nu(\ep,B)=   { |\ep|\over \pi  \hbar^2  \vf^2} \left| 1 + 2
\sum_{{p}=1}^{\infty}     \cos { 2 \pi p  \ep^2 \over {\ep_{B}}^2}  \cos 2 \pi p \gamma \right|  \label{nug}  \ee
After a double integration, we obtain ${\cal N}_0(\ep)$ and $\delta {\cal N}(\ep)$ \cite{SM}. Then, we find that the field dependent part of the grand potential (per unit area) is given by
\be  \delta \Omega_\alpha(\mu,T) =  \\
  r  B^{3/2}  \sum_{p=1}^\infty \frac{\cos 2 \pi p \gamma }{p^{3/2}} \int_{- \infty}^\infty \! \! f'_\mu(\ep) \Delta(2 \sqrt{p}{|\ep| \over \ep_{B}})
  d \ep  \label{GP4} \ee
where $r\equiv { \vf e^{3/2}  \over 2 \pi^2 \sqrt{2 \hbar }}$ and $\Delta(x) \equiv 1 - 2 S(|x|)$ in terms of the Fresnel function $S(x)$. As we now show, the sign of the grand potential depends on the index shift $\gamma$.

\begin{figure}
\begin{center}
\subfigure[]{\includegraphics[width=4cm]{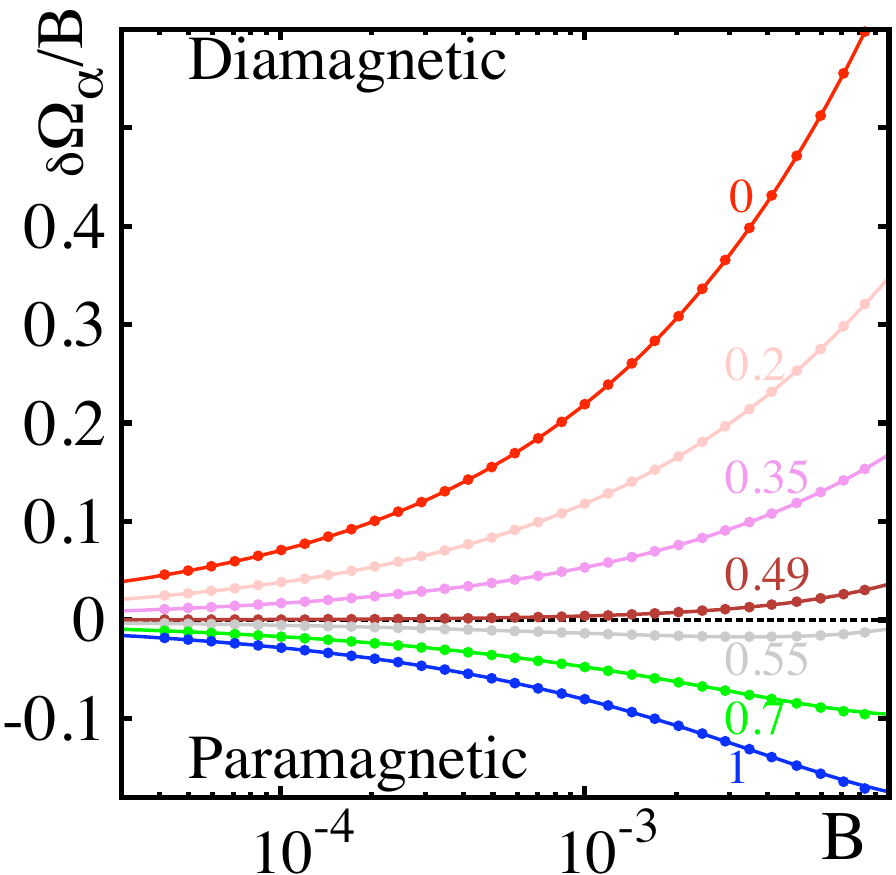}}
\subfigure[]{\includegraphics[width=4cm]{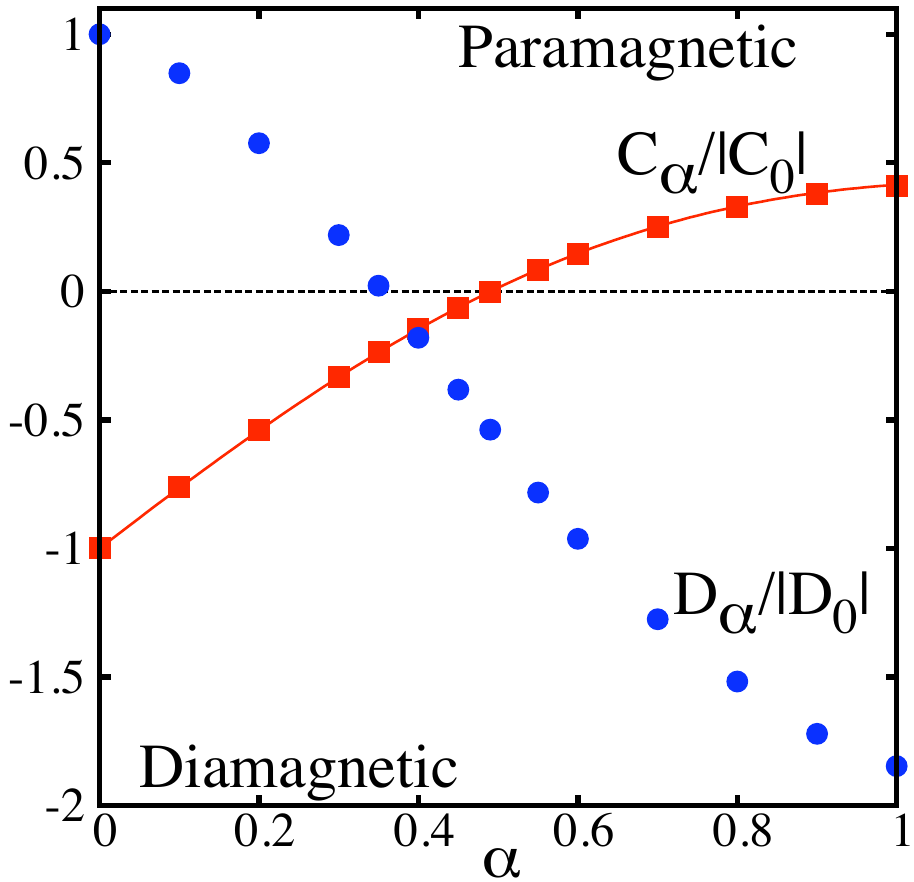}}
\caption{ (a) Field dependent part of the grand potential (per unit area) divided by the magnetic field $\delta \Omega_\alpha/B$ (in units of $et/h$), at zero doping $\mu=0$ and temperature $T=0$, as a function of the magnetic field $B$ (in units of $\frac{4\pi}{\sqrt{3}}\frac{\hbar}{e a^2}$) for various $\alpha$ as indicated. Dots are obtained numerically from the Hofstader spectrum on the corresponding lattice: $\delta \Omega_\alpha/B$ is well fitted by a dependence $-\frac{2}{3}C_\alpha \sqrt{B} -\frac{1}{2} D_\alpha B$, corresponding to a magnetization $\mathcal{M}_\alpha=C_\alpha \sqrt{B}+D_\alpha B$. 
(b) Dimensionless parameters $C_\alpha /|C_0|$ and $D_\alpha /|D_0|$, obtained from the fit, as a function of $\alpha$, with $C_0=-\frac{9\zeta(3/2)}{8\pi^2 \sqrt{2}}\frac{e^{3/2}ta}{\hbar^{3/2}}$ and $D_0\approx 0.094 \frac{e^2 t a^2}{\hbar^2}\approx 4.7|\chi_L|$. Dots are numerical results and the full line is the analytical prediction Eq.~(\ref{eq:M}-\ref{Fngamma}).}
\label{fig:Adealpha}
\end{center}
\end{figure}

First consider the low temperature $T \ll \ep_{B}$ limit. The thermal function $f'_\mu(\ep) \rightarrow - \delta(\ep-\mu)$ so that the grand potential becomes
\be \delta{{\Omega}}_\alpha(\mu,T=0)=  r    B^{3/2}    w_\alpha(\mu / \ep_{\alpha B}), \label{GP2} \ee
where the function $ w_\alpha(x)\equiv   \sum_{p=1}^\infty  {\cos 2 \pi p \gamma \over p^{3/2}}\Delta(2 \sqrt{p} x)$, exhibiting de Haas-van Alphen oscillations, 
is plotted in Fig.~\ref{fig:walpha}b. It generalizes the function calculated by McClure in the case of graphene $\alpha=0$ (Fig. 3 of ref.~\cite{McClure}).
In the particular case $\mu=0$, the magnetization ${\cal M}_\alpha= - {\partial \delta \Omega_\alpha / \partial B}$ is non-analytic (for $\alpha=0$ see \cite{Sharapov})
\be {\cal M}_\alpha  = C_\alpha  \sqrt{B} \quad \mbox{where} \quad  C_\alpha= - (3/2)  \zeta_\gamma(3/2)\, r \label{eq:M} \ee
 and
\be  \zeta_\gamma(n) \equiv  \sum_{p=1}^\infty {1 \over p^n} \cos 2 \pi p \gamma =\mbox{Re}\left[\mbox{Li}_{n}(e^{2 i \gamma \pi})\right] \label{Fngamma}  \ee
with Li$_n(z)$ the polylogarithm function \cite{gradstein}. The square root behavior (\ref{eq:M}) can not be captured by linear response approaches. A similar anomalous scaling $\mathcal{M} \propto \sqrt{B}$ was found for nodal fermions in \cite{Nguyen}. For graphene ($\alpha=0$, $\gamma=0$), the prefactor is $C_0 = - {3  \over 2 } \zeta(3/2)  r= -{3 \vf e^{3/2} \zeta(3/2) \over 4 \pi^2 \sqrt{2 \hbar}} < 0 \label{MM0}$ \cite{Sharapov}; whereas for dice ($\alpha= 1$, $\gamma=1/2$), we find $C_1 = -C_0 (\sqrt{2} -1 ) /\sqrt{2}>0$. Contrary to the case of graphene, the dice lattice is {\it paramagnetic}. This is confirmed numerically (see below) as shown in Fig.~\ref{fig:chi-de-E-alpha}b. More generally, $C_\alpha/|C_0|$ is plotted as a function of $\alpha$ as a full line in Fig.~\ref{fig:Adealpha}b. The magnetization crosses over from dia- to para-magnetism when increasing $\alpha$. The magnetization changes sign when $\alpha_c \simeq 0.49065$, corresponding to $\gamma \simeq 0.19403$. Note that the duality $\alpha \to 1/\alpha$ implies that $\mathcal{M}_{1/\alpha}=\mathcal{M}_\alpha$.

Second, at finite temperature $T \gg \ep_{B}$, on the scale of the thermal function, we can replace $\Delta(x)$ by $(4/\pi) \delta(x)$ so that  Eq.~(\ref{GP4}) shows that the grand potential now varies like $B^2$, which is the standard linear response behavior \cite{AM}. The susceptibility $-\partial^2 \Omega/\partial B^2|_{B\to 0}$ is found to be
\be \chi_\alpha = \chi_0 \frac{\zeta_\gamma(2)}{\zeta(2)} 
\textrm{ where } \chi_0=-\frac{e^2 \vf^2}{12 \pi T} \textrm{sech}^2 (\frac{\beta \mu}{2}) 
\label{chichi0}
\ee
as found by McClure when $\alpha=0$ \cite{McClure} (spin degeneracy is not included here) and where $\zeta_\gamma(2)$ is defined in Eq.~(\ref{Fngamma}). In the $T \rightarrow 0$ limit (still with $T \gg \ep_{B}$), it can be written as $\chi_0=-(e^2 \vf^2 / 3 \pi) \delta(\mu)$. The sign change of $\chi_\alpha$ occurs at $\alpha \simeq 0.51764$, which is not exactly at the same value as for $\mathcal{M}_\alpha$ in the $T\ll \ep_B$ limit. Duality implies that $\chi_{1/\alpha}=\chi_\alpha$.
\begin{figure}
\begin{center}
\includegraphics[width=4cm]{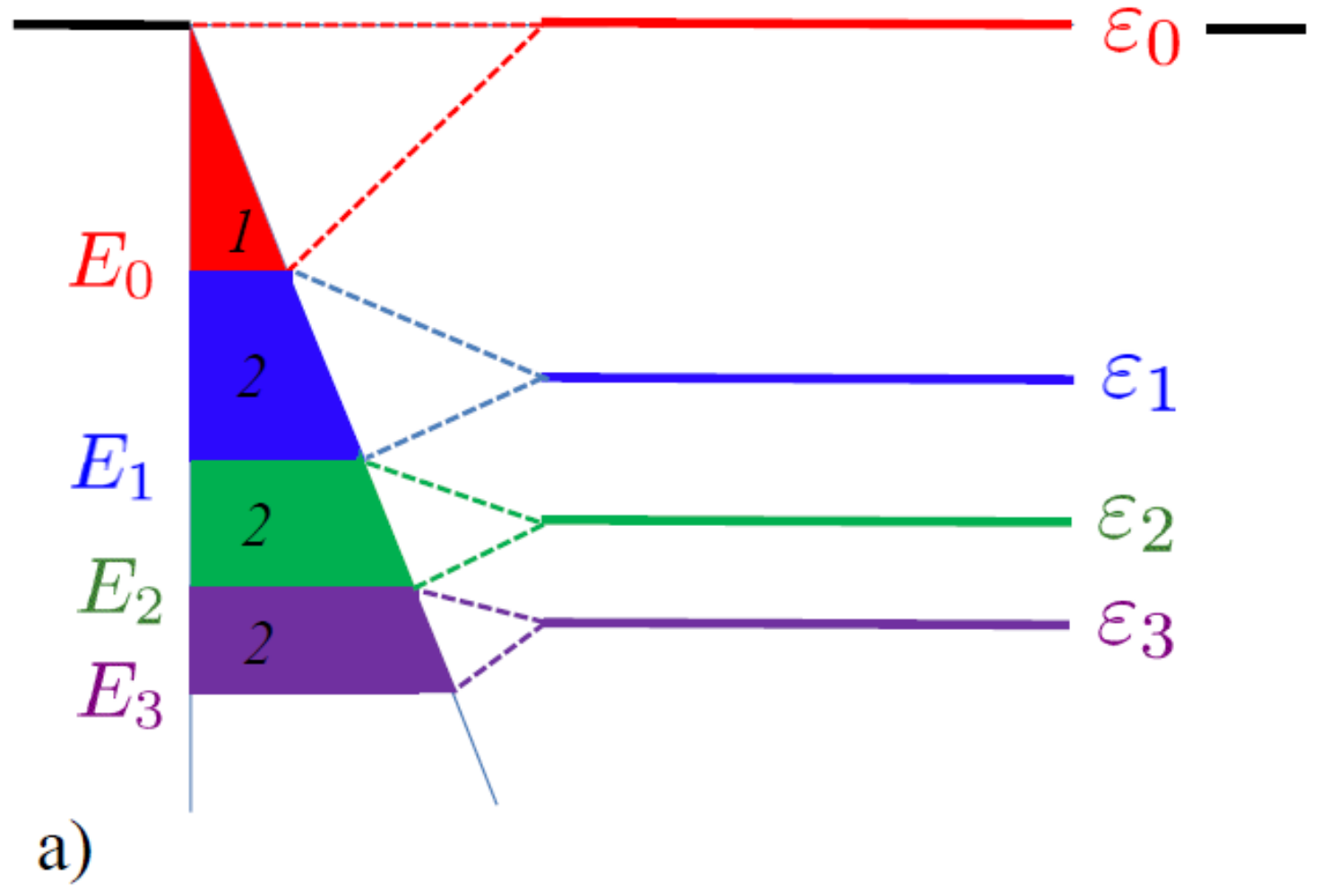}
\includegraphics[width=4cm]{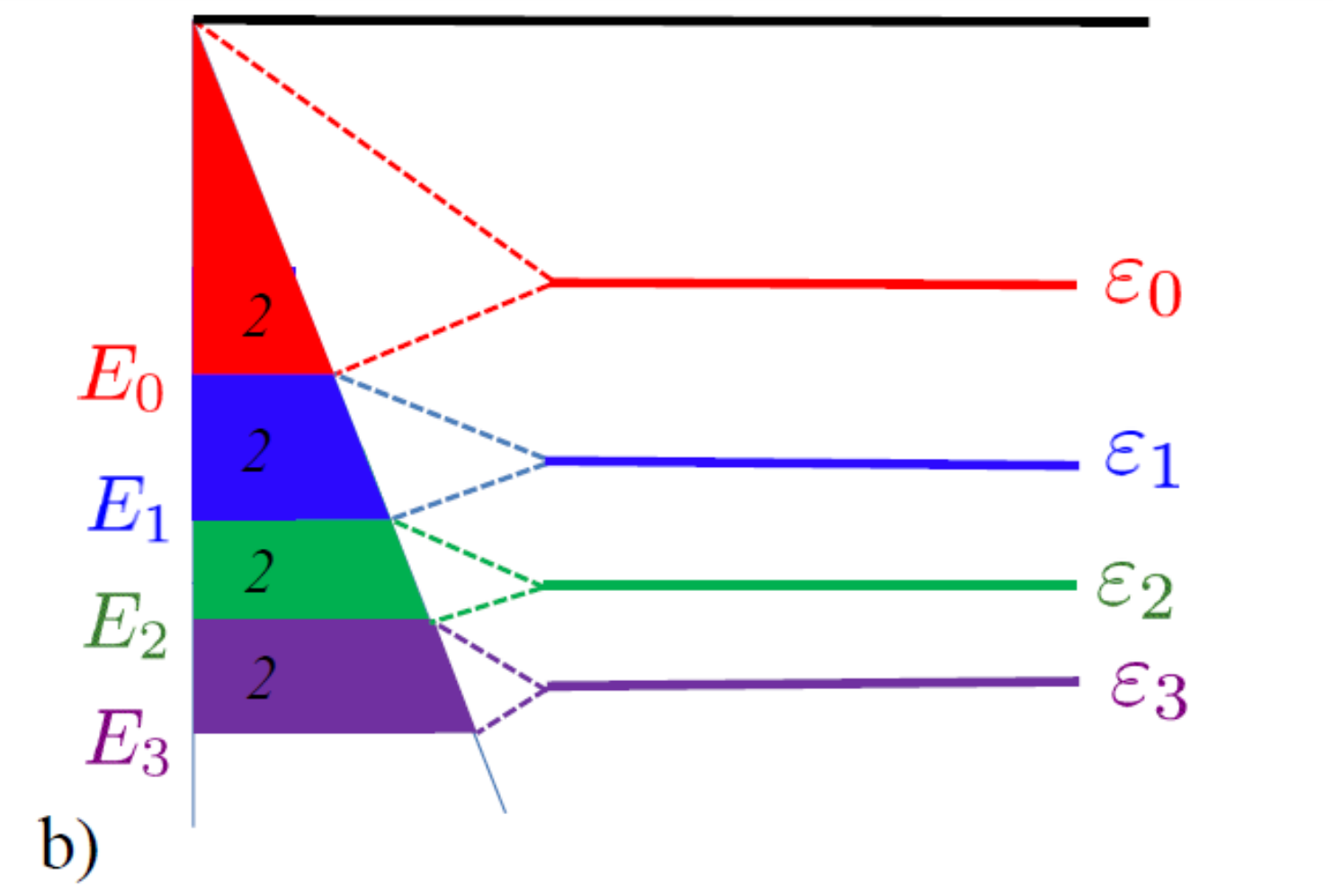}
 \caption{Coalescence of the zero field spectrum (lower band near zero energy) into Landau levels $\ep_n$. (a) Graphene ($\alpha=0$): $\ep_n = - \sqrt{|n|}\,  \ep_{B}$; $E_n= - \sqrt{n+1/2}\, \ep_{B}$ with $n \geq 0$. (b) Dice ($\alpha=1$): $\ep_n = - \sqrt{|n+1/2|}\, \ep_B$; $E_n= - \sqrt{n+1}\, \ep_B$ with $n \geq 0$. The area of  each colored slice $[E_{n-1},E_n]$ counts the total number of states in this slice in zero field. In a field, these states condense into LL.}
 \label{fig:dia-para}
\end{center}
\end{figure}
\medskip

{\it Numerics - } We now consider the energy levels in a magnetic field for the infinite $\alpha$-$\mathcal{T}_3$ lattice (Hofstadter spectrum, see \cite{Vidal} for $\alpha=1$ and \cite{Rammal} for $\alpha=0$), and compute the grand potential numerically. For a similar numerical approach, see \cite{Koshino}. First at $T=0$ and $\mu=0$, we find that the magnetization $\mathcal{M}_\alpha$ is the sum of an anomalous $C_\alpha \sqrt{B}$ and a regular $D_\alpha B$ contribution (Fig.~\ref{fig:Adealpha}a). The coefficients $C_\alpha$ and $D_\alpha$ are plotted in Fig.~\ref{fig:Adealpha}b; $C_\alpha$ is properly given by Eq.~(\ref{eq:M}). The $\alpha$-dependent regular contribution $D_\alpha B$ is expected to come from interband contributions due to lattice effects beyond the Dirac cones approximation. In that respect, a perturbative approach, such as that developed in \cite{Stauber} for $\alpha=0$, might be helpful to quantitatively describe the fitted parameter $D_\alpha$.

Next, we compute the finite temperature ($T\gg \epsilon_B$) susceptibility  $\chi$ as a function of $\mu$ in the whole band for different values of $\alpha$ and for $T\ll t$ (Fig.~\ref{fig:chi-de-E-alpha}a). For $\alpha=0$, aside from the central diamagnetic peak, we recover the $\chi(\mu)$ of \cite{Stauber}. For all $\alpha$, we checked the following sum rule: the integral of the orbital susceptibility over the whole band $\int d\mu \chi(\mu)$ vanishes \cite{SM}. Note that single band approaches \cite{LP}, see also \cite{Fukuyama}, predict an $\alpha$-\textit{independent} susceptibility that does not satisfy the sum rule. The $\alpha$-dependent $\chi(\mu)$ is an indication of the importance of wavefunction dependent interband effects that deserve a more quantitative study using perturbative approaches. At $\mu=0$, the susceptibility is a linear function of $1/T$ (Fig.~\ref{fig:chi-de-E-alpha}b) with a slope which is well fitted by $-e^2v^2\zeta_\gamma(2)/(12\pi\zeta(2))$ as predicted by Eq.~(\ref{chichi0}). The slope changes sign from dia- to paramagnetic behavior at $\alpha \approx 0.52$. 



{\it Discussion - } The physical origin of this dia- to paramagnetic crossover may be understood following usual textbook arguments. Figure~\ref{fig:dia-para} shows how the zero field spectrum coalesces into LLs for the cases of graphene and dice lattices. In the case of graphene, the contribution of each slice $[E_{n-1},E_{n}]$, which condenses into the LL of energy $\ep_n$, decreases the energy, and therefore gives a {\it paramagnetic} contribution. However the contribution of the first (red) slice $[0,E_{0}]$, which condenses into the zero energy LL $\ep_0=0$, increases the energy and thus provides a {\it diamagnetic} contribution which actually compensates the total paramagnetic contribution of all other slices, giving a total {\it diamagnetic} contribution.
In the case of the dice lattice, the contribution of all slices is {\it paramagnetic}, leading to a total {\it paramagnetic} contribution. Therefore the contribution of the zero-energy LL in graphene is essential: although its energy is field independent, its degeneracy is proportional to $B$ and the condensation into this mode {\it increases} the total energy. In the intermediate case, the two-fold degeneracy of the levels is lifted (Fig. \ref{fig:landau-graphene-to-dice}a), the $n=0$ LL acquires a finite energy, therefore its contribution becomes less important. There is a continuous cross-over between the two extreme cases represented in Fig.~\ref{fig:dia-para}. The qualitative argument following this figure can of course be turned into a quantitative calculation of the grand potential,
which reproduces the results obtained above.

{\it Conclusion and experimental proposal - }
We have shown that the orbital susceptibility of a system featuring Dirac cones can be continuously tuned from dia to paramagnetic as a function of a hopping parameter $\alpha$. Such an $\alpha$-$\mathcal{T}_3$ model can be realized experimentally with cold fermionic atoms loaded in an optical lattice. Following the proposal of \cite{Rizzi} for the optical dice lattice ($\alpha=1$), one simply needs to dephase one of the three pairs of laser beams to obtain $\alpha\neq 1$ \cite{SM}. 
Adding an artificial $U(1)$ gauge potential to simulate a perpendicular magnetic field \cite{DalibardRMP}, the internal energy and the entropy (and therefore the free energy) of the trapped Fermi gas could be measured following the techniques of \cite{Thomas}. From the dependence of the free energy on the magnetic field, the sign change of the susceptibility as a function of $\alpha$ could be directly tested. De Haas-van Alphen oscillations could be studied as well \cite{Grenier}.


\begin{acknowledgements}
We acknowledge help from G.M. Tia at an early stage of this work and useful discussions with the mesoscopists in Orsay.
\end{acknowledgements}

\newpage
{
\section*{Supplemental material: Grand potential}
Here we follow McClure (see in particular the appendix A in \cite{McClure}) and consider the grand potential for the low-energy Dirac cone model $\Omega(\mu,T)= \int {\cal N}(\ep) f_\mu'(\ep) d\ep$, where ${\cal N}(\ep)=\int_{-\infty}^\ep d\ep' N(\ep')$ is the doubly integrated density of states (DoS) and $N(\ep)=\int_{-\infty}^\ep d\ep' \nu(\ep')$ is the integrated DoS. Starting from the DoS $\nu(\ep)$, we introduce its integrals $N_0(\ep)= \int_0^\ep \nu(\ep) d\ep$ and ${\cal N}_0(\ep) = \int_0^{\ep} N_0(\ep) d\ep$ calculated from the Dirac point at zero energy. Therefore the integrated DoS calculated from the bottom of the band (the lower cut-off $-W$ introduced in the text) is $N(\ep)=  N_0(\ep) - N_0(-W) =N_0(\ep) + N_0(W)$. Since the total number of states has to be field independent  ($N_0(W)=$ constant), the {\it field dependent part} of the integrated DoS is $\delta N(\ep) = \delta N_0(\ep)$ where $\delta N(\ep)\equiv N(\ep,B)-N(\ep,B=0)$. The field dependent part of the doubly integrated DoS is therefore
\be \delta {\cal N}(\ep) =
  \delta {\cal N}_0(\ep) - \delta{\cal N}_0(-W)  \label{GP1}  \ee 
and that of the grand potential is given by $\delta \Omega= \int  \delta {\cal N}(\ep) f_\mu'(\ep) d\ep$}. 

The quantity
${\cal N}_0(\ep)$ is obtained from (\ref{nug}) and reads
\be  {{\cal  N}}_0(\ep,B)= {  1 \over  \pi \hbar^2  \vf^2}
\left|{1 \over 6} \ep^3 + { \ep_B^3 \over 4 \pi} \sum_{{p}=1}^{\infty}  {\cos(2\pi p \gamma)  \over   p^{3/2}}    S\left(   { 2 \sqrt{  p} |\ep| \over \ep_B}  \right)\right| \label{NN1} \ee
where $S(x)$ is a Fresnel integral :
\be S(x)= \int_0^x \sin {\pi t^2 \over 2 } d t \ee
From Eqs.~(\ref{GP1}) and (\ref{NN1}), we obtain the field dependant part of the grand potential (\ref{GP4}).

\section*{Supplemental material: Susceptibility sum rule}
It seems to be well known that there is a susceptibility sum rule for tight-binding models, see e.g.~\cite{Stauber}. However, as we are not aware of a proof, we provide one here. Consider a tight-binding model with Hamiltonian $H=\sum_j \ep_j |j\rangle\langle j| +\sum_{j\neq l}t_{jl}|j\rangle\langle l|$ in real space, where $\ep_j$ are on-site energies and $t_{jl}$ are hopping amplitudes. In a magnetic field, the hopping amplitudes $t_{jl}$ are multiplied by $\exp(i\frac{e}{\hbar}\int_j^l \mathbf{A}\cdot d\mathbf{l})$, where $\mathbf{A}$ is a vector potential (Peierls substitution), and the on-site energies are unaffected. First, $\textrm{Tr} \mathbb{I}=\sum_j 1$, $\textrm{Tr} H=\sum_j \ep_j$ and $\textrm{Tr} H^2=\sum_{jl}|H_{jl}|^2=\sum_{jl}|t_{jl}|^2$ are field independent. Second, at $T=0$, $\int_{-\mathcal{W}_B}^\mu \Omega(\ep,B)d\ep=\int_{-\mathcal{W}_0}^\mu \Omega(\ep,B)d\ep$, where $\pm \mathcal{W}_B$ is the energy of the band edges in a magnetic field $B$ (choosing the middle of the band at zero energy). The band edge energy $\mathcal{W}$ should not be confused with the high-energy cutoff $W$ defined in the text and which limits the validity of the Dirac cone low-energy description. For example, in the simplest tight-binding model of graphene, $\mathcal{W}_0=3t$, while $W\lesssim t$. Third, $\int_{-\mathcal{W}_0}^{\mathcal{W}_0}\Omega(\ep,B)d\ep=\frac{1}{2}\textrm{Tr} H^2+\mathcal{W}_0\textrm{Tr} H - \frac{\mathcal{W}_0^2}{2} \textrm{Tr} \mathbb{I}=$ constant. Now, the zero temperature susceptibility is given by $\chi(\mu,0)=-\frac{\partial^2 \Omega}{\partial B^2}|_{B\to 0}$ and therefore by differentiating the previous result twice with respect to $B$, we find the sum rule $\int_{-\mathcal{W}_0}^{\mathcal{W}_0} \chi(\mu,0) d\mu=0$. This result can easily be extended to finite temperature and we obtain $\int_{-\infty}^{\infty}\chi(\mu,T) d\mu=0$.

\section*{Supplemental material: $\alpha$-$\mathcal{T}_3$ optical lattice}
\begin{figure}[h]
\begin{center}
\subfigure[]{\includegraphics[width=4cm]{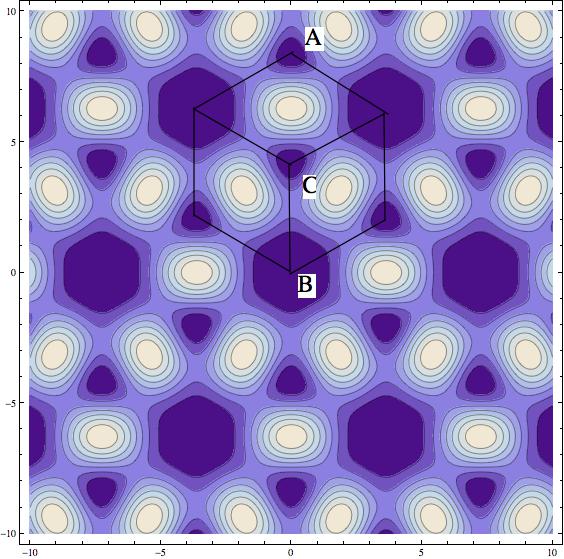}}
\subfigure[]{\includegraphics[width=4cm]{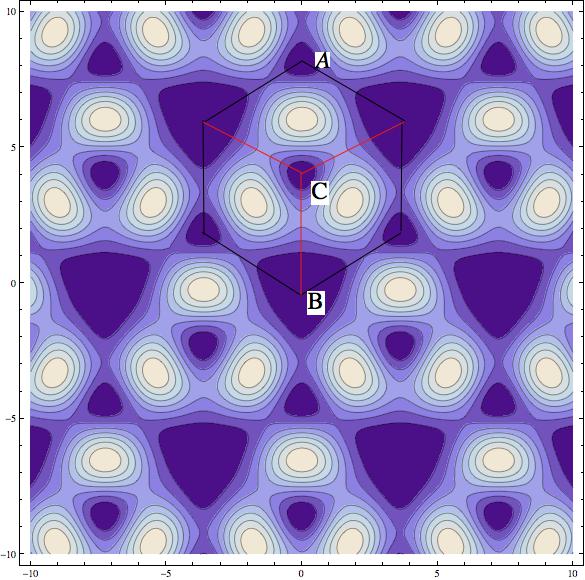}}
 \caption{Contour plot of the $\alpha$-$\mathcal{T}_3$ optical lattice potential $I(x,y)$. (a) No dephasing ($\alpha=1$). (b) Dephasing $\phi=0.3$ ($\alpha<1$).}
 \label{fig:opticallattice}
\end{center}
\end{figure}
We follow the proposal of \cite{Rizzi}. Take three pairs of lasers beams with wavevector $k_L$ making a $2\pi/3$ angle. The electric field is linearly polarized in the $xy$ plane. These three pairs of laser beams interfere to produce an optical potential given by the modulus square of the total electric field. This gives the following dice-like potential
\begin{eqnarray}
I(x,y)&\propto& 3\sin^2(\frac{k_L y}{2}) \sin^2(\frac{k_L x \sqrt{3}}{2})\nonumber \\
&+&[\cos(k_L y) - \cos(\frac{k_L y}{2}) \cos(\frac{k_L x \sqrt{3}}{2})]^2
\end{eqnarray}
shown in Fig.~\ref{fig:opticallattice}a. The on-site energies of the hub ($B$) and rims ($A$ and $C$) are equal. The nearest neighbor hopping amplitudes are such that $t_{AC}\ll t_{AB}=t_{BC}$, corresponding to $\alpha=1$. The inter-site distance is $a=4\pi/(3 k_L)=2 \lambda_L /3$.

Now dephasing one of the three pairs of lasers by $\phi$ gives:
\begin{eqnarray}
I(x,y)&\propto& 3\sin^2(\frac{k_L y}{2}) \sin^2(\frac{k_L x \sqrt{3}}{2})\\
&+&[\cos(k_L y+\phi) - \cos(\frac{k_L y}{2}) \cos(\frac{k_L x \sqrt{3}}{2})]^2 \nonumber
\end{eqnarray}
which realizes the $\alpha$-$\mathcal{T}_3$ model (see Fig.~\ref{fig:opticallattice}b). The parameter $\alpha$ is controlled by the dephasing $\phi$. The on-site energies are still equal and $t_{AC}\ll t_{AB},t_{BC}$, but $t_{AB}\neq t_{BC}$, which means that $\alpha \neq 1$.


\begin{thebibliography}{}
\bibitem{rescaling}After rescaling, the lattice model has hopping parameters $t_{AB}=t/\sqrt{1+\alpha^2}$ and $t_{CB}=\alpha t_{AB}$.
\end{thebibliography}
\end{document}